\documentclass[letter]{aa}
\usepackage[varg]{txfonts}
\usepackage{ulem}
\usepackage{siunitx}
\usepackage{natbib}
\usepackage[thinc]{esdiff}
\bibpunct{(}{)}{;}{a}{}{,} 
\usepackage[colorlinks=true, linkcolor=blue, citecolor=blue, filecolor=blue, urlcolor=blue]{hyperref}

\begin{document}

\title{Strong clumping in global streaming instability simulations with a dusty fluid}

\author{Dominik Ostertag \inst{1,2} \and Mario Flock \inst{1}}

\institute{Max-Planck-Institut für Astronomie, Königstuhl 17, 69117 Heidelberg, Germany, \email{ostertag@mpia.de} \label{inst1} 
\and
Fakultät für Physik und Astronomie, Universität Heidelberg, Im Neuenheimer Feld 226, 69120 Heidelberg, Germany \label{inst2}}

\abstract
{The process of planet formation in protoplanetary disks and the drivers behind the formation of their seeds are still major unknowns. It is a broadly accepted theory that multiple processes can trap dusty material in radially narrow rings or vortex-like structures, preventing the dust from drifting inwards. However, it is still necessary to identify the relevant process behind the clumping of this dusty material, which can result in its collapse under gravity. One promising candidate is the streaming instability arising from the aerodynamic interaction between dust and gas once their densities are similar.}
{We used a global disk model based on recent observational constraints to investigate the capacity of the streaming instability to form dust clumps, which would then undergo gravitational collapse. Furthermore, our goal is to verify the observability of the produced structures using Atacama Large Millimeter/submillimeter Array (ALMA) or Next Generation Very Large Array (ngVLA).}
{For the first time, we present global 2D ($R,z$) hydrodynamic simulations using the FARGO3D code, where the dust is treated as a pressureless fluid. The disk model assumes stratification, realistic boundary conditions, and meaningful resolution to resolve the fast-growing modes. We chose two values for the total dust-to-gas mass ratio $Z=0.01$ and $Z=0.02$. We then compared the maximum clump density to the local Hill density and computed the optical depth of the dust disk.}
{With a dust-to-gas mass ratio of $Z=0.01$, we confirm previous streaming instability simulations, which did not indicate any ability to form strong concentrations of dust clumps.
With $Z=0.02$, dense clumps form within 20 orbits; however, they reached only 30$\%$ of the Hill density, even when applying disk parameters from the massive protoplanetary disks GM Aur, HD163296, IM Lup, MWC 480, and TW Hya, which all share astonishingly similar surface density profiles.}
{Our results show that clumping by the streaming instability to trigger self-gravity is less efficient than previously thought, especially when more realistic density profiles are applied. By extrapolating our results, we estimated the gravitational collapse of concentrated pebbles earliest at 480 orbits; whereas for more frequent, less massive, or more compact disks, this time frame can reach up to 1000 orbits.
Our results predict that substructures caused by streaming instability can vary between optical thin and optical thick at ALMA Band 1 wavelength for less massive disks.
However, the average clump separation is 0.03 au at 10 au distance to the star, which is far too small to be observable with ALMA and even ngVLA.
For the currently observed disks and best-fit surface density profiles, we predict efficient planetesimal formation outside 10 au, where the ratio of Hill- and gas midplane density is sufficiently small. Our results suggest that even for massive Class II disks, the critical Hill density can be reached in dust concentrations within 480 - 1000 orbits, corresponding to tens or hundreds of thousands of years, depending on the radial position.}

\maketitle

\section{Introduction}
To properly study and understand the possible formation of planets and planetary seeds, it is crucial to understand the interplay between the gas and dust components within protoplanetary disks \citep{Armitage2019, Lesur2023}. Small grains, which are well coupled to the gas, follow the gas structure, while larger grains are less coupled and settle towards the midplane. This natural behavior will then result in the accumulation of larger grains and, therefore, an increase in volumetric dust density, making it possible to obtain similar dust and gas densities. Because grains at the midplane are not completely decoupled from the gas, we need to consider the aerodynamic drag effects from gas onto the dust and the impact of dust onto the gas as a result of similar densities. Many works have shown that these drag effects result in so-called drag instabilities \citep[see, e.g.,][]{Squire2017} where the streaming instability \citep{Youdin2004} is one promising representative, which can lead to a natural clumping of dust. This clumping can strongly increase the local dust density compared to the background dust disk \citep[see ][where this clumping was first described for a stratified disk in radial-azimuthal and radial-vertical direction, respectively]{Johansen2007, Johansen2009}. Studying this clumping effect and its dependency on key disk parameters gains insights into the likelihood that streaming or other drag instabilities initiate the formation of planetesimals by the collapse of dense dust accumulations. Many works have investigated the influence of the global dust-to-gas ratio \citep[see also][]{Johansen2009, Yang2014, Li2021}, the gas pressure gradient \citep[see, e.g.][]{Sekiya2018, Abod2019}, and external turbulence by considering additional disk instabilities \citep[see, e.g.][]{Schaefer2020, SchaeferJohansen2022} or by including forced turbulence \citep[see, e.g.][]{Gole2020, Lim2023}. 

One common approach to studying streaming instability is using local shearing box simulations. However, to investigate the effect of streaming instability on planet formation, we need to consider global simulations to give a clump enough time to drift inwards and collect more material. \cite{Flock2021} provided a model to study this instability with more realistic conditions for the gas pressure profile in a global stratified model and a reasonable resolution to resolve the fastest growing modes using Lagrangian dust grains. In addition to modeling dust as individual particles, we can treat it as a pressure-less fluid if the Stokes number, \textit{St}, is below 0.1. We refer to \cite{Weber2019}, \cite{dustyRossbyWaveInst2023}, or \cite{Ziampras2024} for an incomplete list of works that already use dust as a fluid. The work of \cite{Kowalik2013} presented a study of the streaming instability in an unstratified disk, treating dust as a pressureless fluid. To our knowledge, no work exists in which a global stratified protoplanetary disk was investigated, with a focus on streaming instability and its influence on clump formation where the dust is modeled as a fluid. The structure of this work is as follows. Firstly, we describe the methods we used in Sect. \ref{sec:methods}. Then, in Sect. \ref{sec:results}, we simulate two disks with a dust-to-gas mass ratio of 1\% and 2\%, respectively. We end this work with a discussion in Sect. \ref{sec:discussion} and our conclusion in Sect. \ref{sec:conclusion}.
\section{Method and disk model}
\label{sec:methods}
The setup of our disk model is the same as described in \cite{Flock2021} (see equations 1-10 therein)\footnote{
We note that deviations from the initial conditions of the velocity might lead to similar results. We tested the sensitivity of the initial condition by switching the gas and dust velocity and found no significant differences.}. Therefore, our setup is 2D-axisymmetric using spherical coordinates in radial and meridional direction ($r$, $\theta$).
To model dust as a pressureless fluid, we used the publicly available hydrodynamics code FARGO3D \citep[see][]{Benitez-Llambay2016}, namely, the version aimed at studies of one gas and potentially multiple dust species \citep[][]{Benitez-Llambay2019}. This approach includes the effect of aerodynamic drag both from gas onto dust and from dust onto gas (often referred to as backreaction) and the implementation of dust diffusion \citep[see][]{Weber2019}. We note that the last public commit is from July 22, 2020. 
In Appendix \ref{sec:dampingZoneFARGO3D}, we provide a detailed overview of applied damping and refilling techniques for gas and dust. We set the spherical location of the inner (outer) damping zone to $r_{\mathrm{inf}} = 9.349$ au ($r_{\mathrm{sup}}=10.643$ au). In the inner and outer damping zones, we damp the gas azimuthal and radial velocity; in the outer zone, we damp the vertical dust velocity. We also dampened the gas density in the entire domain back to the initial condition. Each damping happens on a dimensionless timescale of $T = 0.0316$. For the sake of brevity, we provide an overview of the boundary conditions in Appendix \ref{sec:BoundaryConditions}.

\begin{table}[h]
\caption{Initial setup values to study streaming instability in 2D}
\label{table:1}
\centering
\begin{tabular}{l l | l l}
\hline\hline
Parameter & Value & Parameter & Value\\
\hline
$\theta_{\mathrm{min}}$, $\theta_{\mathrm{max}}$    &  1.556796, 1.584796                 &  $N_{\mathrm{\theta}}$                    &  1024          \\
$r_{\mathrm{min}}$, $r_{\mathrm{max}}$              &  9.3 au, 10.7 au                    &  $N_{\mathrm{r}}$                         &  4096          \\
$\Sigma_0$                                          & $\qty{390}{g.cm^{-2}}$              &  $M_{*}$                                  &  $M_{\odot}$   \\
$\Sigma_{\mathrm{d},0}$                             & $\qty{3.9}{g.cm^{-2}}$  ($Z=0.01$)  &  \textit{St}                            &  0.01          \\
$\Sigma_{\mathrm{d},0}$                             & $\qty{7.8}{g.cm^{-2}}$  ($Z=0.02$)  &  $\Pi$                                    &  0.07          \\
$T$                                                 & $2 \pi /\Omega_{\mathrm{K}}(R=1 \mathrm{au})$   &  $\eta$                       &  0.0049        \\  
$\tau_{SD}$                                         &   $2 \pi /\Omega_{\mathrm{K}}(R=1 \mathrm{au})$ &                               &                \\
\hline
\end{tabular}
\tablefoot{The gas surface density $\Sigma_{\mathrm{0}}$ was chosen such that the resulting midplane density at 10 au agrees well with the midplane density profiles in Fig. \ref{fig:densityProfileComparison}. We determine the initial value of $\eta$ and $\Pi$ with equation 5 in \cite{YoudinJohansen2007} and equation 3 in \cite{Li2021}, respectively. The values for $R_0$, $H_{0}/R$, $H_{\mathrm{d},0}/R$, $P$, $Q$ are the same as in Table 1 in \cite{Flock2021}.}
\end{table}

\section{Results}
\label{sec:results}

\begin{figure*}[h]
    \centering
    \includegraphics[]{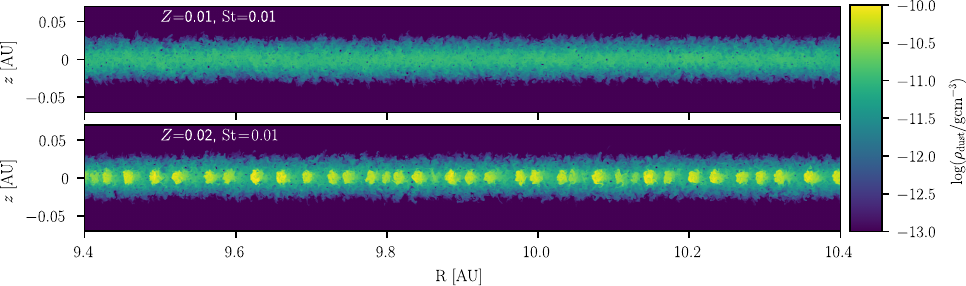}
    \caption{Comparison between a dust fluid simulation with a metalicity of $Z$=0.01 (top) and $Z=0.02$ (bottom) after 100 orbits at 10 au.}
    \label{fig:dustDensity2DComparisonZ001Z002}
\end{figure*}

One feature of streaming instability is strong clumping, which allows the disk to form clumps with higher density than the background \citep{Yang2017, Li2021, Lesur2023}. To be relevant for planet formation processes, these clumps must show densities around the Hill density, $\rho_{\mathrm{Hill}}$, to collapse gravitationally. This section will investigate the ability to form these clumps for two surface density ratios ($Z=\Sigma_{\mathrm{d}}/\Sigma_{\mathrm{g}}=0.01$, $Z=0.02$). An overview of all relevant disk parameters is given in Table \ref{table:1}.
Figure \ref{fig:dustDensity2DComparisonZ001Z002} shows the difference between two simulations after 100 orbits, each starting with one of the values mentioned above. We can directly see that the higher surface density ratio leads to separated clumps.

\subsection{Clump evolution with $Z=0.02$}
\label{sec:clumpingZ002}

\begin{figure}[]
    \centering
    \includegraphics[width=\columnwidth]{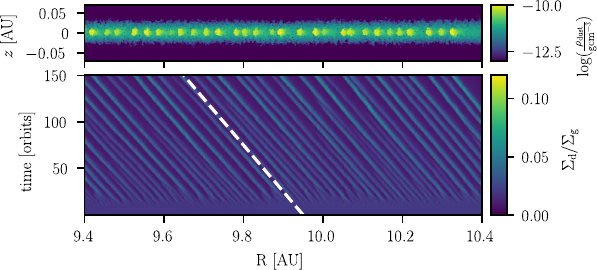}
    \caption{Evolution of the disk with $\Sigma_{\mathrm{d}}/\Sigma_{\mathrm{g}}=0.02$. Top: 2D dust density after 160 orbits, bottom: Evolution of the surface density ratio over 160 orbits. The white dashed line shows the theoretical radial drift with a constant drift velocity of $v_{\mathrm{r, clump}}$ = \qty{30}{cm.s^{-1}}.}
    \label{fig:EvolutionDTGRatio_Z002}
\end{figure}

The disk with $Z=0.02$ is evolved for 160 orbits to see the clump drifting behavior. In Fig. \ref{fig:EvolutionDTGRatio_Z002}, we show the evolution of the surface density ratio $\Sigma_{\mathrm{d}} / \Sigma_{\mathrm{g}}$ as a function of radial position and time as well as the 2D dust density at 160 orbits. 

We indicate the theoretical radial drift due to a constant drift velocity of $v_{\mathrm{r, clump}}= \qty{-30}{cm.s^{-1}}$ with a white line that we fitted by eye. In addition, we compare this clump drift velocity to the drift velocity of the background dust disk, $v_{\mathrm{r, bg}}$, which we determined with Eq. 10 in \cite{Birnstiel2024}, following the solution by \cite{Nakagawa1986}. We determined $\epsilon_{\mathrm{bg}} \approx 0.5$ and $v_{\mathrm{K}} - v_{\mathrm{gas},\varphi} = \eta \cdot v_{\varphi, \mathrm{K}} \approx \qty{4680}{cm.s^{-1}} $ in the midplane at a position of \qty{9.72}{\astronomicalunit} after 100 orbits, which yields a value of $v_{\mathrm{r, bg}} \approx \qty{-41}{cm.s^{-1}}$. This comparison shows that due to the higher dust-to-gas ratio inside the formed clumps (up to $\epsilon \approx 6$), they drift inwards slower than the background dust disk.

\begin{figure}[h]
    \centering
    \includegraphics[width=\columnwidth]{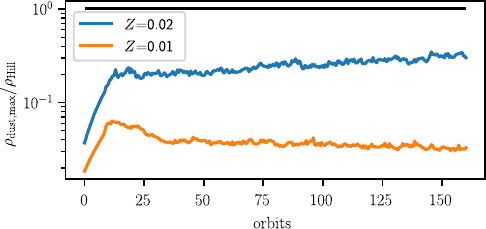}
    \caption{Evolution of the maximum dust density between 9.4 and 10.4 au over 160 orbits for the considered simulations. The black line indicates the density ratio clumps must reach to collapse under their gravity. The chosen domain is meant not to exclude clumps that are an artifact from the outer boundary.}
    \label{fig:evolutionMaxDustDensity}
\end{figure}

The result for a higher ratio of surface densities by just a factor of 2 shows that secondary clumping appears very early in the disk's evolution and the entire domain. This evolution agrees with \cite{Li2021} (see their Fig. 2, lower panel), despite their use of Lagrangian particles to model dust and local shearing box simulations. The question is whether the clump densities are high enough to be relevant for gravitational collapse by reaching the Hill density,
\begin{equation}
    \rho_{\mathrm{Hill}} = \frac{9}{4 \pi} \frac{M_{*}}{R^3},
\label{eqn:hillDensity}
\end{equation}
with $R$ the cylindrical distance to the star and $M_{*}$ the mass of the central star \citep[see appendix B of][also for the difference between Roche- and Hill density]{Klahr2020}. To determine whether the clump density is high enough, we plot in Fig. \ref{fig:evolutionMaxDustDensity} the maximum density for the simulation with $Z=0.02$ between 9.4 au and 10.4 au and compare it with the evolution of maximum dust density for the simulation with $Z$=0.01. We include a horizontal line at $\rho_{\mathrm{dust,max}} / \rho_{\mathrm{Hill}} = 1$.

The comparison indicates that with a dust-to-gas density ratio of $Z=0.02$, the maximum density due to secondary clumping after 160 orbits is roughly 30\% of the Hill density at 10 au. In addition, the evolution of the maximum dust density shows an exponential behavior over time once the streaming instability sets in. The maximum density is related to many clumps that alternate in having the highest density, and extrapolating this exponential behavior, reaching the Hill density would take roughly 480 orbits. To compare our results with those of local shearing box simulations, for instance, the works of \cite{Li2021} or, more recently \cite{Lim2024}, we need to consider that our pressure gradient value is 0.07 (instead of the typically used value of 0.05). \cite{Li2021} have provided (via their Eq. 9) a method to convert from $Z_{\mathrm{crit}}(\textit{St}=0.01,\Pi=0.05) \approx 0.016$ to $Z_{\mathrm{crit}}(\textit{St}=0.01, \Pi=0.07) \approx 0.022$. Because this value is slightly larger than our chosen value, we would not expect to be in the strong clumping region of the parameter space in \cite{Li2021} (from their Fig. 1). In \cite{Lim2024}, the value of $Z_{\mathrm{crit}}(\textit{St}=0.01, \Pi=0.05)$ is reduced to values $<$ 0.01 due to higher resolution. Even with our pressure gradient, this would lead to $Z_{\mathrm{crit}}(\textit{St}=0.01, \Pi=0.07)$ $<$ 0.014, which is fulfilled in our simulation. Hence, with most recent local simulations running for much longer timescales, we expect to be in the strong collapsing part of their parameter space.

Investigating the spatial separation of clumps from Fig. \ref{fig:dustDensity2DComparisonZ001Z002} and \ref{fig:EvolutionDTGRatio_Z002} indicates that 33 clumps form within one au, leading to a separation of roughly 0.03 au. The gas scale height at 10 au is 0.7 au, resulting in a mean clump separation of $H_{\mathrm{g}}$/23.

\subsection{Comparison of disk midplane gas densities to critical Hill density}
\label{sec:densityComparison}

In the previous section, we mentioned that the critical Hill density for gravitational collapse only depends on the star's mass and the distance from the star, showing a substantial decrease with this distance. In contrast, the midplane gas density profiles can have different radial behaviors depending on the power-law index and the density at the characteristic scaling radius. These variations naturally occur due to various processes that dominate the disk evolution \citep[][]{miotello_2022}. In this section, we want to compare five different midplane gas density profiles to their respective Hill density to see if and where sweet spots for planetesimal formation exist; namely, where the ratio of Hill- to midplane density becomes sufficiently small. 

\begin{figure}
    \centering
    \includegraphics[]{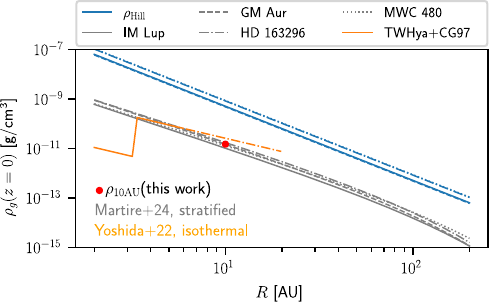}
    \caption{Comparison between five midplane gas density profiles (grey, orange) and their critical Hill densities (blue). For reference, the red dot indicates our work's central location and gas density. A similar plot (albeit with only one gas density profile) was already shown in \cite{Klahr2020}.}
    \label{fig:densityProfileComparison}
\end{figure}

First, we use the recent constraints of disk parameters in \cite{Martire2024} for the four disks MWC 480, IM Lup, GM Aur, and HD 163296 by using the values from their stratified models. For each disk, we can compute the profile of the gas surface density $\Sigma_{\mathrm{g}}$, the midplane gas scale height $H_{\mathrm{g}}$, the midplane gas density using $\rho_{\mathrm{g}} = \Sigma_{\mathrm{g}} / \sqrt{2 \pi} H_{\mathrm{g}}$, and the Hill density. The fifth midplane density is determined from the surface density profile of TW Hya, given in \cite{yoshida_2022}, in combination with the scale height from \cite{Chiang1997}
\begin{equation}
    H_{\mathrm{CG97}} = 0.045 \cdot R \cdot \left( \frac{R}{au} \right)^{2/7},
    \label{eqn:ChiangGoldreich1997}
\end{equation}
which serves as an approximation for this almost face-on disk.
The same gas scale height profile is used to determine the radial profile of the sound speed using $c_{\mathrm{s}} = H_{\mathrm{CG97}} \cdot \Omega_{\mathrm{K}}$ for TW Hya. We then approximate the gas angular velocity with the Keplerian angular velocity. The solid line indicates the radial domain in which the surface density profile was robustly determined. \cite{yoshida_2022} noted the fact that the real profile beyond five au must be smaller than the extrapolation to avoid becoming gravitationally unstable. For reference, we added the midplane gas density of our model at a location of 10 au. The comparison of these profiles to the Hill density is shown in Fig. \ref{fig:densityProfileComparison}.

Firstly, it should be pointed out that the profiles for four disks from \cite{Martire2024} define a narrow band that agrees well with the profile of TW Hya for radii smaller than six au. Secondly, the Hill-to-midplane-gas-density ratio decreases when the radius increases to roughly 50-60 au. For the disk IM Lup, the density ratio decreases from 50 at 10 au to 30 at 30 au. This suggests that the critical density is reached more easily outside 10 au.

\section{Discussion}
\label{sec:discussion}

A massive disk with $Z=0.02$, $\Sigma_{\mathrm{g}}(10\mathrm{au})=390$ $\mathrm{g/cm^2}$, and $H_{\mathrm{g}}=0.7$ au already produces denser dust concentrations after approximately 20 orbits. However, even after 160 orbits of evolution, these clumps only show a density of 30 $\%$ of the Hill density. In total, 480 orbits would be necessary to reach this critical density by continuous growth. This suggests that the first gravitationally unstable clumps form early in the disk's evolution. An important point to consider here is that the four considered disks from \cite{Martire2024} are known as some of the most luminous disks observed so far and may also be the heaviest. Less heavy disks will be more frequent, raising the question of how much time clumps need in these disks to become as dense as the Hill density. Rescaling our isothermal disk surface density with a factor of 6.5 to make it comparable to the disk in \cite{Flock2021} and extrapolating the maximum dust density shows that at least 1000 orbits are necessary to form dense enough clumps. By comparing the evolution time of 1000 orbits to the one from simulation Z2t1 in \cite{Li2021} (noting that we assume their evolution time to be 3000 $t\Omega \approx 477$ orbits to allow clumps to reach the Hill density, see also their Fig. 2 fourth panel), we would expect to see an increase by a factor of roughly 2.

It should be mentioned that no theory has predicted yet how the maximum local dust density evolves in time due to non-linear streaming instability. This is especially true for the strong clumping phase, where multiple clumps compete for the highest dust density. This means we can use no analytic function to extrapolate the profile from Fig. \ref{fig:evolutionMaxDustDensity}. Our choice of an exponential profile is based on the fact that once the non-linear state of the instability sets in, producing the first filaments, the profile shows a linear behavior in the linear logarithmic plot. Hence, there is no guarantee that the profile will continue to be of exponential shape. Still, it might represent the merging of filaments and existing clumps becoming denser with time.

On the other hand, when these clumps move inwards and grow in density, the radial drift becomes smaller as the dust shields itself from aerodynamic drag, creating a traffic jam. This traffic jam allows the clumps to collect additional mass from larger radii. The distance of one au (approximately two au) traveled by the clumps in a massive (less massive or compact) disk after 480 (1000) orbits could then be seen as upper limits. Additional simulations with larger radial domains could be performed to verify the necessary number of orbits for reaching the Hill density and the small radial drift during this time; however, this is beyond the scope of the present work.  

In Sect. \ref{sec:clumpingZ002}, we explain how more dust-enriched clumps move inwards with a smaller velocity than the background dust disk. We expect these clumps to drift inwards even slower as they force the gas to rotate with Keplerian velocity when they grow in dust-to-gas ratio. However, the question of why clumps do not slow down more significantly will be an open question for future projects.

A comparison of midplane density profiles for five disks with the Hill density (see Fig. \ref{fig:densityProfileComparison}) shows that the critical density tends to decrease faster with $R$ than the disk profiles. This suggests that a sweet spot for dense enough clumps forming due to streaming instability might exist in the outer part of the disk where the critical density is closer to the gas midplane values. However, these sweet spots may be too far from the star (e.g., 50 au), therefore, such orbital timescales would be too large and clump formation would take too long to be relevant. 

In Sect. \ref{sec:clumpingZ002}, we mention the fact that the separation of formed clumps at 10 au is much smaller than one au; in addition, Fig. \ref{fig:clumpOpticalDepth_Z002} suggests that for a less massive or compact disk, only Band 1 of the Atacama Large Millimeter/submillimeter Array (ALMA) would be capable of offering a good contrast between the optically thick clumps and the optically thin background. Because the resolution of ALMA is typically on the order of a few au, we would not be able to resolve the clump separation. The upcoming Next Generation Very Large Array (ngVLA) will also observe at long wavelengths (ALMA Band 3 and longer wavelength) with an estimated increase in the baseline by an approximate factor of 10 \citep[see][]{Ricci2018}. Even with a resolution of 1 au, the distance between the clumps will be impossible to resolve. However, we note that \cite{Scardoni2024} has introduced a method to detect unresolved dust rings in protoplanetary disks using azimuthal brightness variations of inclined disks. 
resolve
\section{Conclusion}
\label{sec:conclusion}
For the first time, we present global dust-fluid models of streaming instability in stratified protoplanetary disks that undergo a strong clumping phase. Using realistic density and temperature profiles, streaming instability sets in for a disk with a dust-to-gas ratio of $Z=0.02$, with the first clumps already observed after 20-25 orbits, confirming previous works.
Investigating the maximum dust density of these clumps indicates that 30$\%$ of the Hill density can be reached after 160 orbits when considering massive disks such as GM Aur, IM Lup, HD 163296, MWC 480, or TW Hya, which share astonishingly similar surface density profiles. Extrapolation suggests the maximum density to reach the critical Hill density within 480 orbits, whereby clumps drift only by a maximum of 1 au. For less massive or compact disks, we find an increase by a factor of 2 in the necessary number of orbits and the distance that clumps drift inwards. We highlight that depending on the midplane dust density profile, the Hill density usually decreases faster than the fluid densities. This could make it possible to reach clump densities close to the critical density faster in the outer parts of the disk. Even though such clumps might already exist in observed protoplanetary disks, we could not resolve them with ALMA or the ngVLA due to their small spatial separation.

\begin{acknowledgements}
We thank Francesco Zagaria for the helpful discussion about recent constraints of disk parameters that led to the final version of Fig. \ref{fig:densityProfileComparison}. In addition, we want to thank the referee for the supportive report and suggestions for improving this work.
\end{acknowledgements}

\bibliographystyle{bibtex/aa.bst} 
\bibliography{bibtex/reference}

\begin{appendix}

\section{Boundary conditions}
\label{sec:BoundaryConditions}

With respect to boundary conditions for gas and dust, we proceeded as follows: In the meridional direction, we use the central difference scheme to approximate the gas density gradient and used a simple forward Euler to extrapolate the density into the ghost cells. We did not allow gas inflow into the domain at the same boundaries by setting the ghost cell values to 0 in case the velocity points into the domain. Using the same method, we also did not allow the inflow of gaseous material into the domain at the radial boundaries. Gas was only allowed to leave the domain. We used symmetrical boundary conditions for those components of the gas density and 3D gas velocity that are not explicitly specified, meaning that the ghost cell value is a copy of its corresponding value in the computational domain. For the dust density and the dust velocity components, we used symmetric boundary conditions in the radial and meridional directions, but preventing dusty material from entering the domain in both dimensions using the same method described for gas.

\section{Damping and refilling in FARGO3D}
\label{sec:dampingZoneFARGO3D}

Compared to the setup from \cite{Flock2021}, we use a different equation to damp gas and dust variables close to the radial boundary conditions. The method is described in \cite{Val-Borro2006}. The equation that describes the damping of a variable \textit{x}($r$,$\theta$, $t$) (e.g., the density of gas and dust and its velocity components) is
\begin{equation}
    \diff{x}{t} = - \frac{x - x_0}{\tau_{\mathrm{SD}}} \cdot W(r),
    \label{eqn:dampingDGL}
\end{equation}
where $x_0$=$x$($r$, $\theta$, t=0) is the value at the initial conditions, $\tau_{\mathrm{SD}}$ defines the timescale on which damping happens, and $W(r)$ defines a parabolic function which equals 1 at the boundary between ghost cells and computational domain (in which damping happens) and which equals 0 at the boundary between not damped and damped region within the computational domain. The numerical solution to Eq. \ref{eqn:dampingDGL} is gained implicitly:
\begin{equation}
    \diff{x}{t} \approx \frac{x^{n+1} - x^{n}}{\Delta t} = - (x^{n+1} - x_0) \cdot \frac{W(r)}{\tau_{\mathrm{SD}}}.
    \label{eqn:implicitApproachDampingDGL}
\end{equation}
Rearranging for $x^{n+1}$ and using $\tau_d = \frac{\tau_{\mathrm{SD}}}{W(r)}$
\begin{equation}
    x^{n+1} = \frac{x^{n} \cdot \tau_d + x_0 \cdot \mathrm{d}t}{\tau_d + \Delta t}.
    \label{eqn:solutionNumericalDamping}
\end{equation}
For FARGO3D the expression of $W(r)$ is different for the inner damping-zone (in the following $W_{\mathrm{in}}(r)$) and outer damping-zone (in the following $W_{\mathrm{out}}(r))$
\begin{equation}
    W_{\mathrm{in}}(r) = \left(\frac{r - r_{\mathrm{sup}}}{r_{\mathrm{max}} - r_{\mathrm{sup}}}\right)^2, \quad W_{\mathrm{out}}(r) = \left(\frac{r_{\mathrm{inf}} - r}{r_{\mathrm{inf}} - r_{\mathrm{min}}}\right)^2,
\end{equation}
where $r_{\mathrm{inf}}$ ($r_{\mathrm{sup}}$) is the spherical radius that defines the location of the inner (outer) damping zone. In \cite{Flock2021}, a dimensionless value of $T=1.0$ was applied for the damping timescale where the characteristic radius of the disk is 1 au. However, we used a characteristic radius of 10 au, but we sought to use the same timescale for damping. This results in a dimensionless timescale of $\tau_{\mathrm{SD}} = 0.0316$

Because dust will drift inwards due to the interaction with the gas, but dusty material would not be allowed to enter the computational domain from larger radii, we used a similar artificial dust refilling method as described in \cite{Flock2021} to add dust in the outer damping zone. In every time step, the dust density of each damping zone cell, $\rho_{\mathrm{d}}$, was compared to the initial value $\rho_{\mathrm{d,0}}$. If the density was found to be smaller than 25\% of the initial value, we reset the density back to its initial value, namely, $\rho_{\mathrm{d}} = \rho_{\mathrm{d,0}}$.

To prevent dust from accumulating at the inner damping zone, we used the same method as in \cite{Flock2021} and reduced the azimuthal dust velocity in every time step with $v_{\mathrm{d, \phi}} = (1-10^{-8}) \cdot v_{\mathrm{d, \phi}}$. This slows down the dust in the azimuthal direction and increases its radial velocity.

In \cite{Flock2021}, an additional gas damping was applied in the entire domain to account for the fact that dust drags gas outside the domain, but gas cannot flow into the domain. No gas damping would result in a continuous reduction of the gas mass and affect dust dynamics. We determined the new value of the gas density $\rho(r,\theta)$ for each grid cell after a time step $\Delta t$ via
\begin{equation}
    \rho(r,\theta) = \rho_0(r,\theta) + \left[\rho((r,\theta),t) - \rho_0(r,\theta)\right] \cdot \mathrm{e}^{2\Delta t / T}.
\end{equation}
We also used a dimensionless timescale of $T = 0.0316$ for this damping technique.

\section{Observability of forming clumps}

\subsection{Dust opacities at ALMA band wavelengths}
\label{sec:dustOpac}

In Sect. \ref{sec:results}, we have set our focus on five disks that are among the most luminous and probably the most massive disks. Even though these disks are easier to observe, they do not represent the typical disk population and their masses. In this section, we set our focus to a more frequent disk type. As discussed in \cite{miotello_2022}, a significant fraction of observed disks may consist of very compact disks with critical radii smaller than 15 au and high surface density in the inner region. In addition, they argue that no ALMA sample exists that allows entanglement between a radially extended, low-mass disk or a radially compact disk. Both scenarios allow us to consider a disk that shows a smaller surface density at 10 au than the value in Tab. \ref{table:1}. We will use the value specified in \cite{Flock2021}, which can be obtained from our disk by a rescaling factor of 6.5. 

\begin{figure}[h]
    \centering
    \includegraphics[]{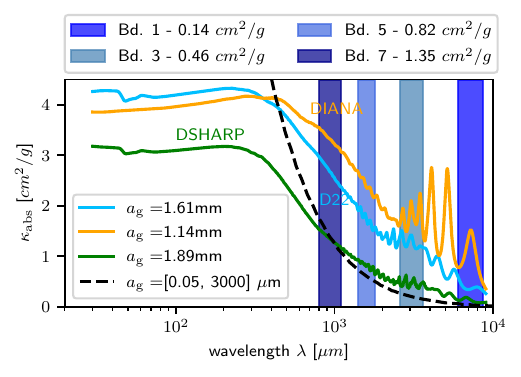}
    \caption{Dust absorption opacities using the dust grain compositions DSHARP (\cite{Birnstiel2018}, DIANA \cite{Woitke2016} and the composition mentioned in \cite{Dullemond2022} (refered to as D22 in the plot). The black dashed line shows the absorption opacity for the default opTool grain size distribution, and each grain consists purely of pyroxene. The vertical bars indicate the wavelength range of different ALMA bands. The mean absorption opacity for each band is calculated based on the DSHARP composition.}
    \label{fig:dustOpacities}
\end{figure}

To identify one or more suitable bands for observing formed clumps, we proceed as follows. We use three different dust grain composition models: The model described in \cite{Dullemond2022} (further referred to as D22), the DSHARP model described in \cite{Birnstiel2018}, and the DIANA model from \cite{Woitke2016}. For all models, we then use the publicly available software opTool \citep{Dominik2021} to compute the dust opacity for absorption in a wavelength range of 0.05 - 9 mm. Because the material density due to the specified compositions varies, we receive different dust grain sizes at 10 au (see legend of Fig. \ref{fig:dustOpacities} for the resulting grain sizes), given a Stokes number of 0.01. Using the wavelength range of each ALMA band, we can compute the mean absorption opacity for each band. We also plot in black the dust absorption opacity for the default grain size distribution of opTool between 0.05 and 3000 $\mu$m, assuming a power-law index of -3.5 and a grain composition purely by pyroxene.

\subsection{Optical depth profile of forming clumps}
\label{sec:ClumpOpticalDepth}
\begin{figure}[h]
    \centering
    \includegraphics[]{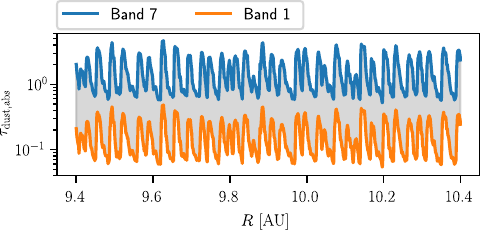}
    \caption{Dust optical depth after 100 orbits as a function of radius for the simulation with $Z=0.02$. The disk we consider here is gained by rescaling the density from the simulation in Fig. \ref{fig:EvolutionDTGRatio_Z002} by a factor of 6.5. The grey region between both curves indicates the possible values of optical depth that are relevant for observations between Band 1 and Band 7.}
    \label{fig:clumpOpticalDepth_Z002}
\end{figure}

The remaining question is whether these clumps can be observed in protoplanetary disks. The ideal scenario is to use, for instance, ALMA with a band in which the clumps are optically thick, but the dust disk around the clumps is optically thin. For further analysis, we stick to the DSHARP model from  \cite{Birnstiel2018} as it leads to the smallest opacities compared to the DIANA model \citep{Woitke2016} and the model described in \cite{Dullemond2022} (See Fig. \ref{fig:dustOpacities} for comparing the three models. To show the range in optical depth for suitable ALMA bands (Band 1 to 7) we use the mean absorption opacity for Band 1 ($\kappa_{\mathrm{Band 1}}$ = \qty{0.14}{cm^2.g^{-1}}) and Band 7 ($\kappa_{\mathrm{Band 7}}$ = \qty{1.35}{cm^2.g^{-1}}), multiply these values with the dust surface density of the considered disk and show the results in Fig. \ref{fig:clumpOpticalDepth_Z002}. The area between both optical depth profiles, relevant for intermediate ALMA bands, is grey. Using the profile from Band 1 shows that the clumps typically have values around 0.3-0.5, while the regions around the clumps are below 0.1. The effect of streaming instability to reduce the surface density of the background disk and, therefore, its optical depth was already mentioned in \cite{Scardoni2021}. They showed that when the streaming instability leads to optically thick clumps, the simulated fluxes in mm wavelength range are consistent with those observed from actual disks.

\end{appendix}

\end{document}